\documentclass[12pt]{article}

\usepackage{tikz}
\usetikzlibrary{decorations.pathreplacing,intersections}
\usetikzlibrary{calc}
\usepackage[margin=1in]{geometry} 
\usepackage{amsmath,amsthm,amssymb}
\usepackage{tikz}
\usepackage{natbib}
\usepackage{subcaption}
\usepackage[utf8]{inputenc}
\usepackage{enumitem}
\usepackage{setspace}
\usepackage{color}
\usepackage{soul}

\newtheorem{prop}{Proposition}[section]
\newtheorem{thm}{Theorem}[section]

\usepackage{cancel}
\newtheorem{lem}{Lemma}[section]
\theoremstyle{definition}
\newtheorem{dfn}{Definition}[section]

\usepackage{ragged2e}

\title{An Alternative Approach for Nonparametric Analysis of Random Utility Models \footnote{I am thankful to the editor Faruk Gul as well as an anonymous referee for helpful comments which greatly improved the paper. I am also thankful to Peter Caradonna, Christopher P Chambers, Federico Echenique, Yuichi Kitamura, Yusufcan Masatlioglu, Alexandre Poirier, Koji Shirai, Mu Zhang, and seminar participants at EC 24, NASMES 2024, and NBER/NSF/CEME 2023 for helpful discussions and comments during the course of this project.\\
Department of Decision Sciences and IGIER, Universit\`{a} Bocconi, E-mail:  \texttt{christopher.turansick@unibocconi.it
}}}

\author{Christopher Turansick}
\date{\today}

\begin{document}

\maketitle

\begin{abstract}
We readdress the problem of nonparametric statistical testing of random utility models proposed in \cite{kitamura2018nonparametric}. Although their test is elegant, it is subject to computational constraints which leaves execution of the test infeasible in many applications. We note that much of the computational burden in Kitamura and Stoye's test is due to their test defining a polyhedral cone through its vertices rather than its faces. We propose an alternative but equivalent hypothesis test for random utility models. This test relies on a series of equality and inequality constraints which defines the faces of the corresponding polyhedral cone. Building on our testing procedure, we develop a novel axiomatization of the random utility model.
\end{abstract}


\section{Introduction}
The random utility paradigm is ubiquitous in modern economics. It is often used to model the choices of a population of rational agents or the repeated choices of a single agent with varying preferences. While the random utility model was initially characterized by \cite{falmagne1978representation} and \cite{mcfadden1990stochastic}, until recently, there has been little work successfully taking these characterizations to real data and testing the random utility hypothesis. \cite{kitamura2018nonparametric} develop an elegant statistical test of the random utility model, allowing for fully nonparametric testing of unrestricted heterogeneity. However, their test is computationally burdensome and quickly becomes infeasible as the number of available alternatives grows. \cite{kitamura2018nonparametric} make note of this computational issue. In response to this, \cite{smeulders2021nonparametric} prove that the testing procedure of \cite{kitamura2018nonparametric} is NP-hard and develop computational tools that vastly reduce the time needed to execute the test of \cite{kitamura2018nonparametric}.

In this paper, we take an alternative approach to that of \cite{smeulders2021nonparametric}. While our goal is the same in that we aim to reduce the computational burden of testing random utility, we instead develop a novel hypothesis test for random utility and show that it offers large improvements over the method of \cite{kitamura2018nonparametric}. To motivate the difference between these two tests, we first note that the set of data points consistent with the random utility model can be written as the convex hull of the data points consistent with deterministically rational choice. This convex set can be represented as either the convex combination of each of these rational choice profiles or it can be represented as the intersection of a finite set of half-spaces. The test of \cite{kitamura2018nonparametric} utilizes the representation through rational choice profiles while our test utilizes the half-space representation of random utility.

While the half-space representation for random utility is known when the analyst observes choice on every available menu of alternatives \citep{falmagne1978representation}, in general the half-space representation is not known when the analyst observes choice on an arbitrary collection of menus. Our methodology circumvents this issue by using the half-space representation of random utility when every menu is observed. If the data we observe is consistent with random utility, then it admits an extension to data on every menu that is also consistent with random utility. Our methodology introduces slack variables which guarantees that our observed data extends to a full dataset which is consistent with random utility. From a theoretical perspective, we can easily and naively introduce slack variables in order to get a linear program which characterizes random utility when choice data is limited. However, not every choice of slack variable will be amenable to the current set of econometric tools. A key component of our methodology is that we are able to introduce slack variables which lead to linear programs that are amenable to the econometric tools developed in \cite{fang2023inference}. Thus our methodology is not only theoretically implementable, but also implementable from an econometric perspective.

In addition to offering computational improvements, our methodology naturally leads to a novel axiomatization of the random utility model when there are unobserved menus. Our new axiom is a statement about when a local form of feasibility extends to a global form of feasibility. These two types of feasibility concern themselves with the assignment of mass to events of the form ``$x$ is chosen from set $A$" and the capacity of each set $A$ to contain the mass of these events. Our axiom improves over current axiomatizations of random utility on limited domains as it can be stated without reference to the model. The standard axiom is from \cite{mcfadden1990stochastic} and can be recovered by applying the Theorem of the Alternative to the vertex representation (i.e. in terms of deterministic choice functions) of random utility. Our axiom can be recovered by applying the Theorem of the Alternative to the half-space representation of the random utility model.

The rest of this paper is organized as follows. In Section \ref{setup} we formally introduce the random utility model as well as discuss the methodologies of \cite{kitamura2018nonparametric} and \cite{smeulders2021nonparametric}. In Section \ref{methodology} we introduce and develop our testing methodology. In Section \ref{axiomatics} we present our new axiomatization of the random utility model. Finally, in Section \ref{conclusion} we conclude and offer a discussion of the related literature.

\section{Random Utility and Testing}\label{setup}
Our focus is on the abstract discrete choice setup. This differs from the initial setup of \cite{kitamura2018nonparametric} (henceforth KS) who focus on random choice from linear price-wealth budgets. Using the results of \cite{mcfadden2005revealed}, KS show that the testing process in their environment can be reduced to testing a specific version of the abstract discrete choice setup. As such, our focus on the abstract setup is without loss. We will later show how to encode properties such as monotonicity into the abstract setup.

\subsection{Model}
Let $X$ be a finite set of alternatives with typical elements denoted $x,y,$ and $z$. We assume that an analyst observes choice on some arbitrary collection of subsets of $X$. We let $\mathcal{X} \subseteq 2^X \setminus \{\emptyset\}$ denote this collection of subsets with typical subsets denoted as $A$ and $B$.\footnote{$2^X$ denotes the power set, the collection of each subset, of $X$.} Throughout we will assume that agents have strict preferences over $X$.\footnote{This assumption can be done away with while keeping with our methodology if the analyst has access to a stronger form of data than what we assume here (see \cite{barbera1986falmagne} and \cite{gul2013random}).} In the setup of KS, this is equivalent to assuming single-valued demand. With this assumption in mind, let $\succ$ denote a linear order over $X$ and $\mathcal{L}(X)$ denote the set of linear orders over $X$. Our analyst has access to data in the form of a random choice rule.

\begin{dfn}
    A function $p:X \times \mathcal{X} \rightarrow \mathbb{R}$ is a \textbf{random choice rule} if it satisfies the following.
    \begin{itemize}
        \item $p(x,A) \geq 0$ for all $x \in A$
        \item $\sum_{x \in A} p(x,A)=1$ for all $A \in \mathcal{X}$
    \end{itemize}
\end{dfn}

In the setup of KS, a random choice rule represents the aggregate choices from a population of agents. Alternatively, a random choice rule can represent the choices of a single agent aggregated across time. The random utility model supposes that there is some distribution over preferences which induces our observed random choice rule. Let $\nu \in \Delta(\mathcal{L}(X))$ denote a typical probability distribution over linear orders of $X$.

\begin{dfn}
    A random choice rule $p$ is \textbf{stochastically rationalizable} if there exists a probability distribution over linear orders $\nu$ such that the following holds for all $A \in \mathcal{X}$ and $x \in A$.
    \begin{equation}
        p(x,A) = \sum_{\succ \in \mathcal{L}(X)}\nu(\succ) \mathbf{1}\{x \succ y \text{ } \forall y \in A \setminus \{x\}\}
    \end{equation}
\end{dfn}

\subsection{Current Methodology}
We now discuss the current methodology for testing the random utility model. Our focus is on the theory and computational burden of the test of KS rather than the statistical properties. Further, our discussion will restrict to the case of idealized data. That is to say, we assume that $p(x,A)$ is the true choice frequency of $x$ from $A$. In reality, an analyst would observe some number of choices from the choice set $A$ and $\hat{p}(x,A)$ would be subject to finite sampling error. For a discussion of the statistical properties as well as implementation with real data, we turn the reader to \cite{kitamura2018nonparametric} and \cite{smeulders2021nonparametric}. 

\subsubsection{A Conic Approach}
To best understand the methodology of KS, we first rewrite the definition of stochastic rationality in matrix form. Consider a matrix $M$ whose rows are indexed by $\mathcal{L}(X)$, the set of linear orders of $X$, and whose columns are indexed by pairs of the form $(x,A)$, where $x \in A \in \mathcal{X}$. The element $m_{\succ,(x,A)}=1$ if $x$ is the maximal element of $A$ according to $\succ$ and $m_{\succ,(x,A)}=0$ otherwise. Now suppose we can encode a probability distribution over preferences as a vector $\nu$ whose indices agree with the rows of $M$ and our random choice rule as a vector $p$ whose indices agree with the columns of $M$. By doing so, the definition of stochastic rationality can alternatively be given as follows.

\begin{dfn}
    A random joint choice rule $p$ is \textbf{stochastically rationalizable} if     \begin{equation}\label{storatMatEq}
        \exists \nu \in \Delta(\mathcal{L}(X)) \text{ such that } \nu^T M = p.\footnote{Note that, compared to \cite{kitamura2018nonparametric}, we take the transpose of each matrix.}
    \end{equation}
\end{dfn}

The first observation that leads to the test of KS is that we can relax the assumption that $\nu$ is a probability distribution. Specifically, we need only assume that $\nu \geq 0$. This is because each row in $M$ encodes a (rational) choice function and $\sum_{x \in A} p(x,A)=1$ for both choice functions and random choice rules. Formally, this means we can rewrite Equation \ref{storatMatEq} as
\begin{equation}\label{conicLinEq}
    \exists \nu \geq 0 \text{ such that } \nu^T M = p.
\end{equation}
The second observation that leads to the test of KS is that we can turn this existence problem into a quadratic minimization problem. Formally, for a positive definite matrix $\Omega$, there exists a solution to Equation $\ref{conicLinEq}$ if and only if
\begin{equation}
    \min_{\nu \geq 0} (\nu^T M - p)^T \Omega (\nu^T M - p)= 0.
\end{equation}
The point of transforming the original definition of stochastic rationality into this quadratic form is that the test statistic and the bootstrap technique proposed by KS relies on working with conic shape constraints in this quadratic form.

A key insight from KS related to Equation \ref{conicLinEq} is that the random utility model, as well as other convex models of stochastic choice, can actually be represented as cones in Euclidean space. From Equation \ref{conicLinEq}, it follows that testing if our data is stochastically rationalizable is equivalent to checking if the data is contained by the following cone.
\begin{equation}\label{PCone}
    \mathcal{P}=\{p|p=\sum_{\succ \in \mathcal{L}(X)} \nu(\succ) m_{\succ}, \nu(\succ) \geq 0\} 
\end{equation}
Equation \ref{PCone} is known as the vertex representation or V-representation of a polyhedral cone. The V-representation of a cone simply says that a cone is every point that can be generated as a convex combination of each extremal ray of the cone. Each finitely generated cone has an alternative half-space representation or H-representation. The H-representation simply says that a finitely generated cone can always be represented as the intersection of finitely many half-spaces. The equivalence between these two representations is due to the Weyl-Minkowski Theorem.

\begin{thm}[Weyl-Minkowski Theorem]\label{weylmink}
    A subset $\mathcal{P}$ of $\mathbb{R}^H$ is a finitely generated cone
    \begin{equation}
        \mathcal{P}=\{\sum_{i=1}^H m_i \nu_i | \nu_i \geq 0 \} \text{ for some } M=[m_1, \dots, m_H] \in \mathbb{R}^{I \times H}
    \end{equation}
    if and only if it is a finite intersection of closed half-spaces
    \begin{equation}
        \mathcal{P}=\{p \in \mathbb{R}^H | Np \geq 0 \} \text{ for some } N \in \mathbb{R}^{H \times J}.
    \end{equation}
\end{thm}

In general, when the H-representation is known, the testing procedure for random utility can be simplified. One could apply tools from the econometric literature on moment inequalities (see \cite{andrews2010inference}, \cite{bugni2010bootstrap}, \cite{canay2010inference}, and \cite{cox2023simple}). However, the H-representation of random utility varies with $\mathcal{X}$ and the exact forms of many of these representations are still unknown (see \cite{gilboa1990necessary}, \cite{gilboa1992game}, and \cite{cohen1990random}). Notably, the H-representation of random utility is known when $\mathcal{X}=2^X \setminus \{\emptyset\}$ and is due to \cite{falmagne1978representation}. This H-representation is important for our methodology and we will discuss it when we introduce our methodology.

For a moment, consider a general convex model of stochastic choice. If this model forms a simplex, then the $M$ and $N$ matrices in Theorem \ref{weylmink} can be chosen to be the same size. This is because simplices have the same number of vertices as they have sides (generating half-spaces). Further, when we are dealing with three or more dimensions, simplices are the only convex shapes with the same number of vertices and generating half-spaces. Another property of simplices is that every point in a simplex can be written as a \textit{unique} combination of its vertices. This means that if our convex model of stochastic choice is unidentified, then the $M$ and $N$ matrices in Theorem \ref{weylmink} can be of different sizes thus leading to potential computational improvements by considering the H-representation of our model. Since the random utility model is known to be unidentified \citep{fishburn1998stochastic,turansick2022identification} and most of the computation time in the testing procedure of KS comes from the construction of $M$, it turns out we can improve on the computation time of KS by working with the H-representation of the random utility model.

\subsubsection{Column (or Row) Generation}

An alternative way to reduce the computation time of KS is presented in \cite{smeulders2021nonparametric}. This method of \cite{smeulders2021nonparametric} is called column generation. In order to better understand column generation, we first make a few observations about the identification problem in random utility. As mentioned prior, the random utility model is unidentified when $|X| \geq 4$. For datasets which fail to have a unique random utility representation, the support of the representation is also unidentified.\footnote{From \cite{turansick2022identification}, it is known that, when a dataset has a random utility representation, the representation is unique if and only if the support of the representation is unique.} For datasets which have a unique random utility representation, when $|X| \geq 4$, these representations necessarily are not full support \citep{turansick2022identification}. This means that, from an ex post perspective, if we consider the $M$ matrix in the KS testing procedure, when $|X|\geq 4$, there will always be rows that are redundant when we find a rationalizing $\nu$. On the other hand, from an ex ante perspective, no single row of $M$ is redundant as our data can always be induced by a degenerate distribution on the rational type that corresponds to any given row. The column generation procedure of \cite{smeulders2021nonparametric} uses the fact that there will always be ex post redundant rational types in the testing procedure of KS.

The column generation procedure begins by guessing that a certain collection of preferences, or rows of $M$, will not be needed in order to rationalize the data. In doing so, the value of $\nu(\succ)$ is set equal to zero for each preference $\succ$ in this collection. Then we construct a matrix $\Bar{M}$ which is the same as matrix $M$ except, for each preference $\succ$ in our guess, it removes the rows associated with $\succ$. This matrix $\Bar{M}$ generates an inner approximation of the cone formed by $M$, so if our random choice rule $p$ lies in the cone formed by $\Bar{M}$, it also lies in the cone formed by $M$. If $p$ does not lie within the cone formed by $\Bar{M}$, then we can add one row to $\Bar{M}$ corresponding to one of the preferences we had assumed to have $\nu(\succ)=0$. We can then check to see if $p$ lies within this new $\Bar{M}$ and repeat. \cite{smeulders2021nonparametric} present a clever pricing problem that allows them to better choose the order which preferences are added back to $\Bar{M}$. Since most of the computation time involved in KS comes from construction of $M$, by iteratively constructing $M$ using the column generation approach, a lot of time is saved in practice.

\section{New Methodology}\label{methodology}
In this section, we discuss an alternative methodology to the one proposed in KS for testing the random utility model. This methodology works with the H-representation of a cone, but does not rely on the analyst knowing the H-representation for every $\mathcal{X}$. Our methodology proceeds in three main steps.
\begin{enumerate}
    \item Find the H-representation of the model when $\mathcal{X}=2^X \setminus \{\emptyset\}$.
    \item Perform a change of variables so that each non-negativity constraint of the H-representation can be represented by non-negativity of a single variable.
    \item Introduce slack variables which guarantee that the random choice rule on $\mathcal{X} \neq 2^X \setminus \{\emptyset\}$ extends to a random choice rule consistent with the H-representation of the model on $2^X \setminus \{\emptyset\}$.
\end{enumerate}
We now highlight the key difference between our methodology and the methodology of KS and its impact on implementability. Our methodology works with the H-representation of a model rather than the V-representation. In doing so, we run into theoretical concerns not present in KS. Typically the V-representation of a model is how we define a convex model of stochastic choice. In the case of the random utility model, each data point that is in the convex hull of classically rational choice rules is consistent with random utility. These classically rational choice rules are the vertices of the random utility model and we know what they look like ex ante. However, given a V-representation, it is either a theoretical or computational exercise to find the H-representation of a model. Taking the computational approach to finding an H-representation defeats the purpose of our methodology, so that leaves us with finding H-representations through theoretical means. This is one disadvantage of our methodology when compared to KS.

We now apply our methodology. Recall that the first step in our methodology is to find the H-representation of random utility on $2^X \setminus \{\emptyset\}$. Luckily, this H-representation is already known and is due to \cite{falmagne1978representation}. Before introducing this representation, we need a bit more notation.

\begin{dfn}
    The \textbf{M\"{o}bius inverse} of a random choice rule $p:X \times 2^X \setminus \{\emptyset\} \rightarrow \mathbb{R}$ is the function $q:X \times 2^X \setminus \{\emptyset\} \rightarrow \mathbb{R}$ which is defined as follows.
    \begin{equation}\label{mobinv}
        \begin{split}
            q(x,A) & = p(x,A) - \sum_{A \subsetneq B}q(x,B) \\
            & = \sum_{A \subseteq B}(-1)^{|B \setminus A|}p(x,B)
        \end{split}
    \end{equation}
\end{dfn}
The second line of Equation \ref{mobinv} was introduced by \cite{block1959random} and is called the Block-Marschak polynomial. In general, the M\"{o}bius inverse $q(x,A)$ captures how much probability is added to or removed from $p(x,\cdot)$ at set $A$. In terms of the random utility model, these M\"{o}bius inverse functions have a strong connection to the probability weight put on contour sets.

\begin{thm}[\cite{falmagne1978representation}]\label{rumidentification}
    A distribution over linear orders $\nu$ is a random utility representation of a random choice rule $p:X \times 2^X \setminus \{\emptyset\} \rightarrow \mathbb{R}$ if and only if the following holds for all nonempty $A \subseteq X$ and $x \in A$.
    \begin{equation}
        q(x,A) = \sum_{\succ \in \mathcal{L}(X)}\nu(\succ)\mathbf{1}\{X\setminus A \succ x \succ A\setminus \{x\}\}
    \end{equation}
\end{thm}

This tells us is that the M\"{o}bius inverse $q(x,A)$ is equal to a probability weight when our random choice rule is stochastically rationalizable. This further means that $q(x,A)$ must be non-negative when our random choice rule is stochastically rationalizable. This turns out to be the H-representation of random utility.
\begin{thm}[\cite{falmagne1978representation}]\label{rumHrep}
    A random choice rule $p:X \times 2^X \setminus \{\emptyset\} \rightarrow \mathbb{R}$ is stochastically rationalizable if and only if $q(x,A) \geq 0$ for all $x \in A \subseteq X$.
\end{thm}
In terms of of the M\"{o}bius inverse, this means that $p$ is stochastically rationalizable if and only if every set $A$ is contributing some non-negative amount to $p(x,\cdot)$. We stated Theorem \ref{rumHrep} in terms of the M\"{o}bius inverse $q$, but recall that, in Equation \ref{mobinv}, the M\"{o}bius inverse is just a linear function of $p$. Thus, by asking that $q(x,A) \geq 0$ for all $x \in A \subseteq X$, we are simply asking that $Np \geq 0$ for some matrix $N$. It should now be apparent that Theorem \ref{rumHrep} gives us the H-representation of random utility when $\mathcal{X} = 2^X \setminus \{\emptyset\}$. It is important to note that Theorem \ref{rumHrep} does not hold regardless of our domain $\mathcal{X}$. As an example, if $\mathcal{X}$ just contains the binary choice sets, having a non-negative M\"{o}bius inverse is without empirical content. This means that we are unable to simply apply Theorem \ref{rumHrep} regardless of our domain.

We now move onto the second and third steps of our methodology. As Theorem \ref{rumHrep} highlights, non-negativity of the M\"{o}bius inverse functions correspond to the H-representation of random utility. Our goal now is to introduce slack variables that guarantee that our random choice rule $p$ on $\mathcal{X}$ extends to $2^X \setminus \{\emptyset\}$ while being consistent with the H-representation of random utility. There are at least two ways we can proceed in introducing slack variables. We can introduce slack variables $\tilde{p}(x,A)$ for sets $A \not \in \mathcal{X}$ that coincide with unobserved choice probabilities. Alternatively, we can introduce slack variables $\tilde{q}(x,A)$ for each set that coincide with the M\"{o}bius inverse of our hypothetical full domain random choice rule. For now, we focus on slack variables that coincide with choice probabilities. Consider the following linear program.
\begin{align}
    \sum_{x \in A}\tilde{p}(x,A) = 1 \text{ }\forall A \in (2^X \setminus\{\emptyset\})\setminus \mathcal{X} \label{psumtoone}\\
    \sum_{A \subseteq B : B \in \mathcal{X}} (-1)^{|B \setminus A|}p(x,B) + \sum_{A \subseteq B : B \not \in \mathcal{X}} (-1)^{|B \setminus A|}\tilde{p}(x,B) \geq 0 \text{ } \forall A \in 2^{X} \setminus \{\emptyset\}, x \in A \label{pblockmars}\\
    \tilde{p}(x,A) \geq 0 \text{ } \forall A \not \in \mathcal{X}, x \in A \label{pnonneg}
\end{align}
By naively using $\tilde{p}$ as our slack variables, we run into a problem. Notably, we have an equality constraint, Equation \ref{psumtoone}, and a non-negativity constraint, Equation \ref{pnonneg}, but we have an additional inequality constraint in Equation \ref{pblockmars}. This additional inequality constraint means that we are unable to directly apply current econometric tools to test the linear program. \cite{fang2023inference} develops a bootstrap technique for testing the hypothesis that there exists some $x$ such that \begin{equation}\label{fangAndsantos}
    Nx=f(p) \text{ for } x \geq 0
\end{equation} where $f(p)$ is some function of the observed data and $N$ is a matrix. Equation \ref{pblockmars} is exactly what prevents us from directly applying the test of \cite{fang2023inference}. Luckily, we can use $\tilde{q}$ as our slack variables to solve this problem.

One problem arises when we move from $\tilde{p}$ to $\tilde{q}$. When using $\tilde{p}$, it is easy to encode that the slack variables induce a random choice rule. Simply ask that $\tilde{p}$ are non-negative and sum to one at every choice set $A$. It is less obvious how to guarantee that $\tilde{q}$ are the M\"{o}bius inverse of a full domain random choice rule. Our next result exactly characterizes when this is the case.
\begin{lem}\label{qtop}\footnote{We also point out that \cite{kono2023axiomatization} contemporaneously developed an analogous result in their Lemma 3.14.}
    A function $q: X \times 2^X \setminus \{\emptyset\} \rightarrow \mathbb{R}$ is the M\"{o}bius inverse of some full domain random choice rule $p$ if and only if it satisfies the following conditions.
    \begin{enumerate}
        \item $\sum_{x \in A}q(x,A) = \sum_{y \in X \setminus A}q(y,A \cup\{y\})$ for all $\emptyset \subsetneq A \subsetneq X$
        \item $\sum_{x \in X}q(x,X)=1$
        \item $\sum_{A \subseteq B}q(x,B) \geq 0$ for all  $x \in A \subseteq X$ 
    \end{enumerate}
\end{lem}
The first two conditions of Lemma \ref{qtop} guarantee that $p$ sums to one at every choice set. The third condition of Lemma \ref{qtop} guarantees that $p$ is non-negative everywhere. With this in mind, we can now apply step two of our methodology using $\tilde{q}$ as slack variables. Consider the following linear program.
\begin{align}
    \sum_{A \subseteq B} \tilde{q}(x,B) = p(x,A)  \text{ } \forall x \in A \in \mathcal{X} \label{qcon}\\
    \sum_{x \in A}\tilde{q}(x,A) = \sum_{y \in X \setminus A} \tilde{q}(y,A\cup\{y\}) \text{ } \forall A \not \in \mathcal{X}, \emptyset \subsetneq A \subsetneq X \label{qinflowoutflow}\\
    \sum_{x \in X} \tilde{q}(x,X)=1 \text{ if } X \not \in \mathcal{X} \label{qinitialcon}\\
    \tilde{q}(x,A) \geq 0  \text{ } \forall x \in A \subseteq X \label{qnonneg}
\end{align}    
Above, Equation \ref{qcon} guarantees that $\tilde{q}$ is the M\"{o}bius inverse of some function that agrees with our random choice rule $p$ on sets we observe. Equations \ref{qinflowoutflow} and \ref{qinitialcon} are just the first two conditions of Lemma \ref{qtop} applied at unobserved choice sets. This guarantees that our extended $p$ function induces choice probabilities on unobserved sets. Equation \ref{qnonneg} plays two roles in that it is the H-representation of random utility on a full domain and it implies the third condition of Lemma \ref{qtop}. The following theorem summarizes our discussion thus far.

\begin{thm}\label{RUMEquivalenceThm}
    The following are equivalent.
    \begin{enumerate}
        \item The random choice rule $p:X \times \mathcal{X} \rightarrow \mathbb{R}$ is stochastically rationalizable.
        \item There exist variables $\tilde{p}(x,A)$ solving Equations \ref{psumtoone}-\ref{pnonneg}.
        \item There exist variables $\tilde{q}(x,A)$ solving Equations \ref{qcon}-\ref{qnonneg}.
    \end{enumerate}
\end{thm}

Once again, we now take a moment to compare our methodology with that of KS. First, when applying Equations \ref{qcon}-\ref{qnonneg} and \cite{fang2023inference} to real data, the only values that need to be estimated are $p(x,A)$ that come from the right hand side of Equation \ref{qcon}. In other words, just as in KS, we only need to estimate choice probabilities; each other variable in our testing procedure is constructed. Second, we can now begin to compare the computational burden of each testing procedure. While the bootstrapping procedure in KS and \cite{fang2023inference} differ, in both KS and our testing procedure, almost all of the computation time comes from construction of a matrix; the $M$ matrix in the case of KS and an $N$ matrix which encodes Equations \ref{qcon}-\ref{qnonneg} in our case.\footnote{Formally, we can represent the existence of variable $q$ satisfying Equations \ref{qcon}-\ref{qnonneg} as the existence of a variable satisfying $Nq=l$ for some matrix $N$ and a vector $l$.} The $M$ matrix of KS has one column for each pair $(x,A)$ with $x \in A \in \mathcal{X}$ and one row for each linear order over $X$. This means that the number of rows in $M$ grows at a rate of $|X|!$. Our $N$ matrix has one column for each pair $(x,A)$ with $x \in A \in 2^{X} \setminus \{\emptyset\}$ and one row for each condition in Equations \ref{qcon}-\ref{qnonneg}.

\begin{prop}\label{growthrate}
    The number of rows in $N$ grows at a rate of $|X|2^{|X|-1}-\sum_{A \not \in \mathcal{X}}(|A|-1)$.
\end{prop}

When $|X| \leq 4$, $M$ has fewer rows than $N$, but when $|X|\geq 5$, $N$ always has fewer rows than $M$. In Table \ref{table:matrixsize} we calculate the number of rows in $M$ and $N$ for a few sizes of $X$ when $\mathcal{X}=2^X \setminus \{\emptyset\}$ as this is the worst case for our $N$ matrix. As Table \ref{table:matrixsize} points out, as $|X|$ grows, the number of rows in $M$ becomes vastly larger than the number of rows in $N$.

\begin{table}[h!]
    \centering
    \begin{tabular}{|c|c|c|}
        \cline{1-3}
        $|X| $& $M$ rows & $N$ rows \\
        \cline{1-3}
        3 & $6$ & $12$ \\
        \cline{1-3}
        4 & $24$ & $32$\\
        \cline{1-3}
        5 & $120$ & $80$\\
        \cline{1-3}
        10 & $3,628,800$ & $5120$ \\
        \cline{1-3}
        15 &  $\sim 1.3\times 10^{12}$ & $245,760$ \\
        \cline{1-3}
    \end{tabular}

    \caption{The number of rows in the $M$ and $N$ matrices are given as a function of $|X|$ under the assumption that $\mathcal{X} = 2^X \setminus \{\emptyset\}$. Notably, in the case of $|X|=15$, the number of rows of $M$ given in this table is an underapproximation.}
    \label{table:matrixsize}
\end{table}

We now take a moment to discuss how the column generation procedure of \cite{smeulders2021nonparametric} can be applied in our methodology. Recall that the column generation procedure begins by guessing that some collection of our choice variables are equal to zero. In our setup, these choice variables correspond to $\tilde{q}$. When $\mathcal{X}=2^X\setminus\{\emptyset\}$, each $\tilde{q}$ is constructed directly from the data, and thus the exact value of each $\tilde{q}$ is directly observable. This means that the column generation procedure offers no improvements in this case. However, when $\mathcal{X}$ is missing sets, there is room to apply the column generation procedure. Notably, when $\mathcal{X}$ is not complete, we are unable to directly construct $\tilde{q}$ from the data. This means that it is possible that some of our slack variables can be chosen to be equal to zero. Thus in this case, the column generation procedure can offer computational improvement when added onto our methodology. However, unlike in the case of the KS testing procedure, there is no ex ante guarantee that we can always guess some $\tilde{q}$ to be equal to zero.

\subsection{Encoding Monotone Choice}
Thus far we have developed a methodology for testing random utility on an abstract discrete choice domain. Until now, we have ignored two components of the problem present in the original work of KS. First, in KS the choice data is not from an abstract discrete choice domain but rather comes from choice frequencies on linear price-wealth budgets. Second, in KS utility functions are restricted to be monotone with respect to the greater than or equal to ordering on $\mathbb{R}^n$. In this section, we discuss how to incorporate and deal with these two problems using our methodology.

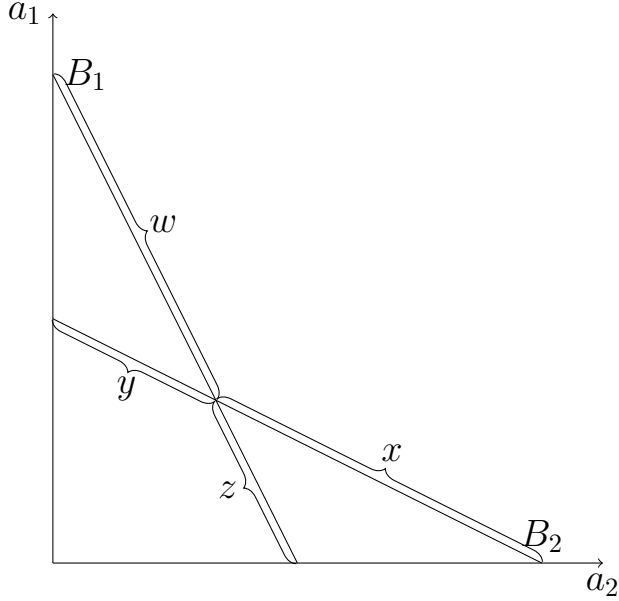
\begin{figure}
    \centering
\begin{tikzpicture}[every node/.append style={font=\scriptsize},scale=3.25]
    
    \path [name path=xaxis,draw,->] (0,0) -- (2.25,0) ;
    \path [name path=yaxis,draw,->] (0,0) -- (0,2.25) ;

    \path[name path=budget1,draw,-] (0,2) -- (1,0);
    \path[name path=budget2,draw,-] (0,1) -- (2,0);

    \draw[decorate,decoration={brace,amplitude=5pt,raise=0.5pt},yshift=0pt] (0,2) -- (2/3,2/3) node [midway,yshift=5pt,xshift=11pt]{\large $w$};
    \draw[decorate,decoration={brace,amplitude=5pt,raise=0.5pt,mirror},yshift=0pt] (2,0) -- (2/3,2/3) node [midway,yshift=11pt,xshift=5pt]{\large $x$};
    \draw[decorate,decoration={brace,amplitude=5pt,raise=0.5pt,mirror},yshift=0pt] (0,1) -- (2/3,2/3) node [midway,yshift=-11pt,xshift=-3pt]{\large $y$};
    \draw[decorate,decoration={brace,amplitude=5pt,raise=0.5pt},yshift=0pt] (1,0) -- (2/3,2/3) node [midway,yshift=-3pt,xshift=-11pt]{\large $z$};

    \node[left] at (0,2.25) {\large $a_1$};
    \node[below] at (2.25,0) {\large $a_2$};
    \node[right] at (0,2) {\large $B_1$};
    \node[above] at (2,0) {\large $B_2$};

\end{tikzpicture}
    \caption{Here we capture two linear budgets over the goods $a_1$ and $a_2$. These budgets are labeled as $B_1$ and $B_2$ and have a single point of intersection. By assuming that no choice occurs at the point of intersection, we are left with $\{w,x,y,z\}$ as the set of patches.}
    \label{fig:linearbudgets}
\end{figure}

We first discuss going between the linear budget domain and abstract discrete choice domain. As is the case in KS and \cite{mcfadden2005revealed}, we will assume we are working with monotone utility functions in the linear budget domain. A result of this monotonicity assumption is that all observed choice should occur on our budget lines. If we observe choice strictly within our budget, then we know that the consumer has money left that they can spend on more of a good to obtain a higher utility. It then follows from an observation made in \cite{mcfadden2005revealed} and used in KS that we need only focus on the partition of our budgets formed by points of intersection. \cite{mcfadden2005revealed} points out that the entire content of random monotone utility with linear budgets is captured by agents' choices on patches. Suppose we have two linear budgets with a single point of intersection, as is the case in Figure \ref{fig:linearbudgets}. Then each of these linear budgets can be partitioned into three components: the point of intersection, ``above" the point of intersection, and ``below" the point of intersection. Each of these three components correspond to a patch in the language of \cite{mcfadden2005revealed} and KS. In Figure \ref{fig:linearbudgets}, if we assume there is no choice at the point of intersection, then the set of patches we are left with is given by $\{w,x,y,z\}$. This set of patches then corresponds to our choice set $X$ in the abstract discrete choice environment. This construction scales beyond the case of two budgets with a single point of intersection.

Our ability to go from the linear budget setup to discrete choice on patches relies on the assumption of monotone utility. Further, the assumption of monotone utility is natural in the linear budget setup of KS. However, the assumption of monotone utility imposes further testable content beyond that of the abstract random utility model. Notably, in the case considered in Figure \ref{fig:linearbudgets}, the H-representation is known. It asks that $p(y,\{x,y\})+p(z,\{w,z\}) \leq 1$. Unfortunately, the H-representation for the linear budget setup becomes increasingly more complicated as the set of budgets and the set of intersections grow. Recall that our methodology works with the H-representation when $\mathcal{X}=2^X \setminus \{\emptyset\}$. This type of data can never be observed in the linear budget setup. That being said, our methodology simply asks if the data we observe can be extended to a full domain H-representation. With this in mind, consider Figure \ref{fig:linearbudgets} and suppose an agent is choosing from the set $\{w,y\}$. The patch $y$ is dominated by $w$ in the standard monotone ordering of $\mathbb{R}^2$. As such, an agent with monotone utility will never choose $y$ when $w$ is available. This observation, initially due to \cite{kashaev2022nonparametric}, turns out to be the only further testable content of monotone random utility beyond that of abstract random utility.

\begin{dfn}
    For $X$ partially ordered by $\rhd$, we say that a random choice rule $p:X \times 2^X \setminus \{\emptyset\} \rightarrow [0,1]$ is \textbf{stochastically rationalizable by monotone utilities} if it is stochastically rationalizable and there exists a rationalizing distribution $\nu$ such that $\nu(\succ)>0$ and $x\rhd y$ imply $x \succ y$.
\end{dfn}

\begin{dfn}
    We say that a random choice rule $p$ is \textbf{monotone} with respect to a partial order $\rhd$ if $x,y \in A$ and $x \rhd y$ implies that $p(y,A)=0$.
\end{dfn}

\begin{thm}[\cite{kashaev2022nonparametric}]\label{RUMmonotone}
    Suppose $X$ is partially ordered by $\rhd$. A random choice rule $p:X \times 2^X \setminus \{\emptyset\} \rightarrow [0,1]$ is stochastically rationalizable by monotone utilities if and only if $q(x,A) \geq 0$ for all $x \in A \subseteq X$ and $p$ is monotone with respect to $\rhd$.
\end{thm}

Theorem \ref{RUMmonotone} gives us the  H-representation we are looking for on a full domain. However, in our Theorem \ref{RUMEquivalenceThm}, we rely on working only with the M\"{o}bius inverse while Theorem \ref{RUMmonotone} utilizes both the base random choice rule and the M\"{o}bius inverse. Our goal now is to transform the monotonicity condition on a random choice rule into a condition on its M\"{o}bius inverse. This turns out to be relatively easy to do given that we already impose that $q(x,A)\geq 0$ in order to get stochastic rationality. For each $x \in X$, let $U(x)$ denote the set of alternatives dominating $x$ according to $\rhd$. Then for each $x$ with nonempty $U(x)$, we can encode monotonicity with respect to $\rhd$ using the following equality constraint.
\begin{equation}\label{partial}
    \sum_{A:x \in A, A \cap U(x) \neq \emptyset}q(x,A)=0 \text{ } \forall x \text{ such that }U(x) \neq \emptyset
\end{equation}
We can then use Equation \ref{partial} along side our prior characterizations of stochastic rationality to get a characterization of stochastic rationality by monotone utilities.
\begin{prop}\label{monotoneequivalence}
    The following are equivalent.
    \begin{enumerate}
        \item A random choice rule $p:X \times \mathcal{X} \rightarrow [0,1]$ is stochastically rationalizable by monotone utilities.
        \item There exists solutions to Equations \ref{qcon}-\ref{qnonneg} and Equation \ref{partial}.
    \end{enumerate}
\end{prop}
Since Equation \ref{partial} is an equality constraint, we are able to use the econometric tools of \cite{fang2023inference} to test for random monotone utility. In terms of the resulting matrix, we only need to add a single row for each alternative $x$ with nonempty $U(x)$. For the growth rate of the number of rows in $N$, the additional monotonicity constraints are of a lower order, being bounded above by $|X|-1$. This means that the rate in Proposition \ref{growthrate} is still approximately correct and is off by at most $|X|-1$ in the case of random monotone utility.

\section{Axiomatics}\label{axiomatics}
Thus far, our focus has been on developing an alternative testing methodology that offers computational improvements over the testing procedure of KS. In this section, we turn our attention to axiomatically characterizing the random utility model using the methodology we developed in the prior section. The main result in this section is a new characterization of random utility on a limited domain. The current standard characterization of random utility on a limited domain is due to \cite{mcfadden1990stochastic}.

\begin{thm}[\cite{mcfadden1990stochastic}]\label{mcfadden}
    A random choice rule $p:X \times \mathcal{X} \rightarrow \mathbb{R}$ is stochastically rationalizable if and only if for any finite sequence $\{(x_i,A_i)\}_{i=1}^n$ with $x_i \in A_i \in \mathcal{X}$ the following holds.
    \begin{equation}\label{ARSP}
        \sum_{i=1}^np(x_i,A_i) \leq \max_{\succ \in \mathcal{L}(X)}\sum_{i=1}^n \mathbf{1}\{x_i \succ y \text{ } \forall y \in A_i \setminus \{x\}\}
    \end{equation}
\end{thm}

We now discuss the relationship between the methodology of KS and Theorem \ref{mcfadden}. As pointed out earlier, the methodology of KS relies on the V-representation of the random utility model. The characterization in Theorem \ref{mcfadden} relies on the V-representation of random utility in the sense that Equation \ref{ARSP} follows from applying the Theorem of the Alternative to the V-representation of random utility (see \cite{borderAlternative} and \cite{border2007introductory} for references). As a result of working with the V-representation and since the V-representation of random utility is itself the model representation of random utility, Equation \ref{ARSP} makes explicit use of the underlying random utility model. That is to say, in order to state Equation \ref{ARSP}, we must make reference to preferences.

Recall that our methodology works with the H-representation of random utility. We get our new axiom by effectively applying the Theorem of the Alternative to the H-representation of random utility. By working with the H-representation, our resulting axiom makes no reference to preferences or the random utility model. Our characterization relies on two supplemental functions.

\begin{dfn}
    A function $c:2^X \rightarrow \mathbb{R}$ is a \textbf{capacity} if $c(\emptyset)=0$.
\end{dfn}

\begin{dfn}
    A function $a:X\times \mathcal{X} \rightarrow \mathbb{R}$ is an \textbf{assignment}.
\end{dfn}

In order to best interpret the role of capacities and assignments, we introduce the following definition.\footnote{Note that our definitions of capacity and assignment differ from those used in cooperative game theory. In cooperative game theory, capacities are typically monotone and assignments typically assign a value to each alternative/agent rather than one value to an alternative for each set containing the alternative.}

\begin{dfn}
    Given a random choice rule $p$, an assignment and capacity pair $(a,c)$ is \textbf{feasible} if the following inequality holds.
    \begin{equation}\label{feasible}
        \sum_{A \in \mathcal{X}}\sum_{x \in A}p(x,A)a(x,A) \leq c(X)
    \end{equation}
\end{dfn}

The role of an assignment $a$ is to assign some mass to the event that $x$ is chosen from $A$. When combined with the probability that $x$ is chosen from $A$, $p(x,A)$, this mass is given by $a(x,A)p(x,A)$. Each set $A$ has a capacity $c(A)$. This capacity must be more than the total mass put on events of the form ``$x$ is chosen from $B$" for each $B \subseteq A$. Feasibility simply asks that the total weight put on choosing some element from some set is less than the capacity of $X$, the total capacity of our environment. Our characterization relies on a second type of feasibility.

\begin{dfn}
    An assignment and capacity pair $(a,c)$ is \textbf{locally feasible} if for each $(x,A)$ with  $x \in A \in 2^X \setminus \{\emptyset\}$ the following inequality holds.
    \begin{equation}\label{localfeas}
        \sum_{x \in B \in \mathcal{X}, B \subseteq A} a(x,B) \leq c(A) - c(A \setminus \{x\}) 
    \end{equation}
\end{dfn}

Local feasibility is a local condition in the sense that it states that the total amount of capacity gained by going from $A\setminus\{x\}$ to $A$ must be more than the total amount of assigned mass to $x$ being chosen in subsets of $A$. Note that, unlike feasibility, local feasibility does not depend on the random choice rule $p$ and the set of locally feasible assignment and capacity pairs can be defined independently of the observed data. Our characterization of stochastic rationalizability is a condition about when local feasibility implies feasibility.

\begin{thm}\label{rumme}
    Given a random choice rule $p:X \times \mathcal{X} \rightarrow \mathbb{R}$, the following are equivalent.
    \begin{itemize}
        \item $p$ is stochastically rationalizable.
        \item Every locally feasible assignment and capacity pair $(a,c)$ is also feasible given $p$.
    \end{itemize}
\end{thm}

Our main innovation over Theorem \ref{mcfadden} is that we use Lemma \ref{qtop} to rewrite the stochastic rationality linear program without reference to any preferences.
\begin{align}
    \sum_{A \subseteq B} q(x,B) = p(x,A) \text{ } \forall x \in A \in \mathcal{X} \label{linconsist}\\
    \sum_{x \in A}q(x,A) = \sum_{z \not \in A} q(z,A\cup\{z\}) \text{ }\forall A \in 2^X \setminus\{X,\emptyset\} \label{lininflow}\\
    \sum_{x \in X} q(x,X)=1 \label{lininitial}\\
    q(x,A) \geq 0  \text{ } \forall x \in A \subseteq X \label{linnonneg}
\end{align}

Equations \ref{linconsist}-\ref{linnonneg} bear an obvious resemblance to Equations \ref{qcon}-\ref{qnonneg}. The notable difference is that we ask that Equation \ref{qinflowoutflow} holds at every choice set rather than simply at unobserved choice sets. In terms of our characterization, this is what gives us that capacity functions are defined on every choice set and the right hand side of Equation \ref{localfeas}. Once we have Equations \ref{linconsist}-\ref{linnonneg}, we simply apply the Theorem of the Alternative and do some minor manipulation in order to get Theorem \ref{rumme}.

\section{Discussion}\label{conclusion}

In this paper we propose a new procedure for testing nonparametric models of random utility. In this test as well as the test of \cite{kitamura2018nonparametric}, much of the computation time is due to calculation of a given matrix. We show that our test will generally lead to a much smaller matrix when compared to the matrix of KS and thus offer large computational improvements over the test of KS.

While our focus in this paper is on testing the random utility model, our testing procedure can be modified for other convex models of choice.\footnote{By convex model of choice, we mean any model of choice where the set of data points consistent with the model is given by a convex set. Frequently it is the case that the model is defined by the extreme points of this convex set.} Recall that our methodology proceeds in three steps.
\begin{enumerate}
    \item Find the H-representation of the model when $\mathcal{X}=2^X \setminus \{\emptyset\}$.
    \item Perform a change of variables so that each non-negativity constraints of the H-representation can be represented by non-negativity of a single variable.
    \item Introduce slack variables which guarantee that the random choice rule on $\mathcal{X} \neq 2^X \setminus \{\emptyset\}$ extends to a random choice rule consistent with the H-representation of the model on $2^X \setminus \{\emptyset\}$.
\end{enumerate}
The first step of this testing procedure extends naturally to any convex model of choice. It becomes a bit more difficult when translating the second and third steps to other convex models. As pointed out by Equations \ref{psumtoone}-\ref{pnonneg} and the discussion thereafter, we are unable to apply the econometric tools of \cite{fang2023inference} if we naively introduce the wrong slack variables in step two of our methodology. The reason why we are able to proceed to step three using Equations \ref{qcon}-\ref{qnonneg} is that Lemma \ref{qtop} characterizes random choice rules in terms of the variables induced by the H-representation of the random utility model. In other words, by the Weyl-Minkowski Theorem, there exists some matrix $H$ such that a random choice rule is consistent with our convex model of choice if and only if $Hp \geq 0$. This $H$ matrix is a linear transformation of choice probabilities. In our case, this $H$ matrix induced our M\"{o}bius inverse function. More generally, this $H$ matrix simply induces some variable $q(H,p)$ which depends on our observed random choice rule as well as the matrix itself. To go between steps two and three, we need to do two things. First, we need to characterize random choice rules in terms of these $q(H,p)$ variables. This characterization will typically consist of equality conditions which correspond to choice probabilities summing to one and inequality conditions which corresponds to probabilities being non-negative. Second, we need to check if non-negativity of our $q(H,p)$ variables implies the non-negativity conditions of the random choice rule characterization. If it does, we are left with a collection of equality constraints and $q(H,p)\geq 0$ as the only set of inequality constraints. The fact that $q(H,p) \geq 0$, or $q(x,A)\geq 0$ in our case, is the only inequality constraint is exactly what lets us use the econometric tools of \cite{fang2023inference}.

\subsection{Related Literature}

Our paper is related to two primary strands of literature. First, our paper is related to the strand of literature started by \cite{kitamura2018nonparametric} which focuses on hypothesis testing the random utility model and other similar models. Following the lead of \cite{kitamura2018nonparametric}, \cite{smeulders2021nonparametric} also studies the problem of hypothesis testing the random utility model and, similar to us, focuses on the problem of computational implementability. However, unlike us, they take the base testing procedure of \cite{kitamura2018nonparametric} as given and try to significantly improve the computation time of that test. In other words, they continue working with the V-representation of the random utility model while we work with the H-representation. In a working paper, \cite{dardanonimethods} studies the bootstrapping procedure used in the hypothesis test of \cite{kitamura2018nonparametric}. The authors propose two alternative bootstrapping techniques and compare the accuracy of these techniques to the one proposed in \cite{kitamura2018nonparametric} using simulations. \cite{fang2023inference} proposes an alternative bootstrap technique which can be used to test the hypothesis of \cite{kitamura2018nonparametric}. There have also been a series of papers using and extending the techniques proposed in \cite{kitamura2018nonparametric} in order to study other related models. \cite{deb2023revealed} extend the techniques of \cite{kitamura2018nonparametric} in order to test a general model of preferences over prices as well as do welfare analysis. \cite{kashaev2022nonparametric} extend the techniques of \cite{kitamura2018nonparametric} in order to test a dynamic version of the random utility model allowing for correlation of preferences over time. Finally, \cite{dean2022better} also use the techniques of \cite{kitamura2018nonparametric} to develop a better test, in the sense of power, for choice overload.

Our paper is also related to the literature which axiomatically studies the random utility model. \cite{falmagne1978representation} is the first to characterize the random utility and does so by asking that the M\"{o}bius inverse of choice probabilities be non-negative. \cite{fiorini2004short} offers an alternative proof of this result using graph theoretic techniques. \cite{monderer1992stochastic} provides an alternate proof of this result using methods from cooperative game theory. \cite{cohen1980random} considers an extension of the result of \cite{falmagne1978representation} to an infinite domain. \cite{nandeibam2009probabilistic} provides a different characterization of random utility using positive linear functionals. \cite{mcfadden1990stochastic} offers a characterization of random utility when the choice domain is incomplete. \cite{stoye2019revealed} offers a short proof of this result using tools from convex analysis. \cite{mcfadden2005revealed} offers an extension of this result to an infinite domain under some regularity conditions. Recently, \cite{gonczarowski2019infinity} extends this result to an infinite domain without any regularity conditions. \cite{clark1996random} offers an alternative characterization of random utility in the case of an incomplete domain using DeFinetti's coherency axiom. \cite{kashaev2022nonparametric} uses techniques developed to study quantum entanglement in order to offer a characterization of dynamic separable random utility on a limited domain. \cite{kono2023axiomatization} study and axiomatize the random utility model when the choice probabilities of a collection of goods are unobservable at every choice set. Lastly, \cite{koida2024dual} studies the the random monotone utility hypothesis in the price-wealth budget domain. They are able to find the H-representation of the model which relies on the ordering of the underlying environment.

\appendix

\section{Preliminary Results}

\subsection{Proof of Lemma \ref{qtop}}
This proof proceeds in two steps. First we show the equivalence between the M\"{o}bius inverse of a function satisfying the first condition in Lemma \ref{qtop} and that function being set constant.
\begin{dfn}
    A function $p:2^X\setminus \{\emptyset\}\rightarrow \mathbb{R}$ is \textbf{set constant} if for all $A,B \in 2^{X} \setminus \{\emptyset\}$ we have that $\sum_{x\in A}p(x,A)=\sum_{y\ in B}p(y,B)$.
\end{dfn}
To note, henceforth, if the M\"{o}bius inverse of a function $p$ satisfies the first condition of Lemma \ref{qtop}, then we will say that the function $p$ satisfies inflow equals outflow. After doing this, the rest of the proof amounts to noting that the second and third conditions are direct translations of $\sum_{x\in X}p(x,X)=1$ and $p(x,A)\geq 0$ into statements about the M\"{o}bius inverse. We begin by showing the equivalence of $p$ satisfying inflow equals outflow and $p$ being set constant. We start with the necessity of inflow equals outflow. Consider a function $f$ with M\"{o}bius inverse $g$ such that $f$ is set constant. We proceed via induction on the size of the complement of $A$. For the base case, let $A=X\setminus\{x\}$. Observe that $f(x,X)=g(x,X)$ We have the following.

\begin{equation*}
    \begin{split}
        \sum_{x \in X A}g(x,A) & = \sum_{x \in A} f(x,A)-g(x,X) \\
        & =\sum_{x\in X}f(x,X) -\sum_{x \in A}f(x,X) \\
        & = f(x,X) =g(x,X)
    \end{split}
\end{equation*}

Above, the first equality holds by the definition of M\"{o}bius inverse. The second equality holds from $f$ being set constant. The third equality follows after collecting like terms. This shows that the base case of inflow equals outflow holds. Now assume that inflow equals outflow holds for all $B$ with $|X \setminus B| < n$. Let $A$ be such that $|X \setminus A|=n$.

\begin{equation*}
    \begin{split}
        \sum_{x \in A} g(x,A) & = \sum_{x\in A}f(x,A) - \sum_{x\in A} \sum_{A \subsetneq A'}g(x,A') \\
        & = \sum_{x\in A}f(x,A) - \sum_{A \subsetneq A'}[\sum_{x\in A'}g(x,A') - \sum_{x\in A' \setminus A}g(x,A')] \\
        & = \sum_{A \subsetneq A'} \sum_{x \in A' \setminus A}g(x,A') - \sum_{A \subsetneq A' \subsetneq X} \sum_{x \in A'} g(x,A') \\
        & = \sum_{A \subsetneq A'} \sum_{x \in A' \setminus A}g(x,A') - \sum_{A \subsetneq A' \subsetneq X} \sum_{y \in X \setminus A'}g(y,A \cup \{y\}) \\
        & = \sum_{z \in X \setminus A} g(z,A \cup\{z\})
    \end{split}
\end{equation*}

Above, the first equality holds by the definition of M\"{o}bius inverse. The second equality just adds zero. The third equality holds as $g(x,X)=f(x,X)$ and because $f$ is set constant. The fourth equality holds by the induction hypothesis. The fifth equality follows from combining like terms. Thus the above string of equalities show that inflow equals outflow is necessary. We now show sufficiency. Now suppose $f$ satisfies inflow equals outflow. Consider some $A \subsetneq X$.

\begin{equation*}
    \begin{split}
        \sum_{x \in A} g(x,A) & = \sum_{x\in A}[f(x,A) - \sum_{A \subsetneq A'}g(x,A')] \\
        & = \sum_{x\in A}f(x,A) - \sum_{A \subsetneq A'}[\sum_{x\in A'}g(x,A') - \sum_{x\in A' \setminus A}g(x,A')] \\
        & =[\sum_{x\in A}f(x,A)-\sum_{x\in X}f(x,X)] + \sum_{A \subsetneq A'} \sum_{x \in A' \setminus A}g(x,A') - \sum_{A \subsetneq A' \subsetneq X} \sum_{x \in A'} g(x,A') \\
        & =[\sum_{x\in A}f(x,A)-\sum_{x\in X}f(x,X)] +  \sum_{A \subsetneq A'} \sum_{x \in A' \setminus A}g(x,A') - \sum_{A \subsetneq A' \subsetneq X} \sum_{y \in X \setminus A'}g(y,A \cup \{y\}) \\
        & =[\sum_{x\in A}f(x,A)-\sum_{x\in X}f(x,X)] +  \sum_{z \in X \setminus A} g(z,A \cup\{z\})
    \end{split}
\end{equation*}
The first equality above holds due to the definition of M\"{o}bius inverse. The second equality just adds zero. The third equality holds as $f(x,X)=g(x,X)$. The fourth equality holds by inflow equals outflow. The fifth equality follows from combining like terms. By inflow equals outflow, we know that $\sum_{x\in A}g(x,A)=\sum_{z \in X \setminus A}g(z,A\cup\{z\})$. This means that the above string of equalities gives us that $\sum_{x\in A}f(x,A)=\sum_{x\in X}f(x,X)$. Since $A$ is arbitrary, this tells us that $f$ is set constant.

Now we return to the case of choice probabilities. Recall that $q(x,X)=p(x,X)$. This means that asking $\sum_{x\in X}p(x,X)=1$ is equivalent to asking that $\sum_{x\in X}q(x,X)$. Further, asking that $p(x,A)\geq 0$ is equivalent to asking that $\sum_{A \subseteq B}q(x,B) \geq 0$ by the definition of the M\"{o}bius inverse. It then follows that asking for the three conditions in Lemma \ref{qtop} to hold is equivalent to asking that the following conditions hold.
\begin{enumerate}
    \item $\sum_{x \in A}p(x,A)=\sum_{y\in B}p(y,B)$ for all $A,B \in 2^{X}\setminus\{\emptyset\}$
    \item $\sum_{x \in X}p(x,X)=1$
    \item $p(x,A) \geq 0$ for all $x \in A \subseteq X$
\end{enumerate}
The above three conditions define a random choice rule. Thus Lemma \ref{qtop} holds.

\subsection{Proof of Lemma \ref{GeneralRUM}}
We now prove a weaker version of Theorem \ref{RUMEquivalenceThm} that will be useful in the proof of Theorem \ref{RUMEquivalenceThm} and \ref{rumme}.
\begin{lem}\label{GeneralRUM}
    The following statements are equivalent.
    \begin{enumerate}
        \item A random choice rule $p:X \times \mathcal{X} \rightarrow [0,1]$ is stochastically rationalizable.
        \item There exists a solution to Equations \ref{linconsist}-\ref{linnonneg}.
    \end{enumerate}
\end{lem}
We now proceed with our proof of Lemma \ref{GeneralRUM}. We proceed with sufficiency of our second condition. Observe the following. Equations \ref{lininflow}-\ref{linnonneg} imply that $\tilde{q}$ is the M\"{o}bius inverse of some full domain random choice rule. Further, Equation \ref{linnonneg} implies that this full domain random choice rule is stochastically rationalizable. Lastly, Equation \ref{linconsist} implies that the full domain random choice rule induced by $\tilde{q}$ agrees with our observed data on sets we observe. Thus our second condition implies stochastic rationalizability. Now we proceed with necessity of the second condition. Observe that if our random choice rule $p$ is stochastically rationalizable, then it admits an extension to a full domain (i.e. $2^{X}\setminus \{\emptyset\}$ that is also stochastically rationalizable. A full domain random choice rule is an extension of our observed random choice rule if and only if its M\"{o}bius inverse satisfies Equation \ref{linconsist}. Further, a full domain random choice rule is stochastically rationalizable if and only if its M\"{o}bius inverse satisfies Equation \ref{linnonneg}. Finally, since Equation \ref{linnonneg} implies the third condition in Lemma \ref{qtop}, if the M\"{o}bius inverse of a function satisfies Equations \ref{lininflow}-\ref{linnonneg} then the function is a random choice rule by Lemma \ref{qtop}. Thus Equations \ref{linconsist}-\ref{linnonneg} are necessary for stochastic rationality.

\section{Omitted Proofs}
\subsection{Proof of Theorem \ref{RUMEquivalenceThm}}
We begin with the equivalence between stochastic rationalizability and Equations \ref{psumtoone}-\ref{pnonneg}. Equations \ref{psumtoone} and \ref{pnonneg} is equivalent to asking that there is a full domain extension of our random choice rule. By Theorem \ref{rumHrep}, we know that a full domain random choice rule is stochastically rationalizable if and only if $\sum_{A\subseteq B}(-1)^{|B \setminus A|}p(x,B) \geq 0$ for all $x \in A \subseteq X$. Thus asking that there exist a solution to Equations \ref{psumtoone}-\ref{pnonneg} is equivalent to asking that there is some full domain extension of our random choice rule that satisfies the conditions of Theorem \ref{rumHrep}. Thus stochastic ratioanlizability is equivalent to there be a solution to Equations \ref{psumtoone}-\ref{pnonneg}.

We now move on to the equivalence between stochastic rationalizability and the existence of a solution to Equations \ref{qcon}-\ref{qnonneg}. We proceed by showing an equivalence between Equations \ref{qcon}-\ref{qnonneg} and Equations \ref{linconsist}-\ref{linnonneg}. As the constraints in Equations \ref{qcon}-\ref{qnonneg} are a subset of the constraints in Equations \ref{linconsist}-\ref{linnonneg}, it follows immediately that if Equations \ref{linconsist}-\ref{linnonneg} have a solution then Equations \ref{qcon}-\ref{qnonneg} have a solution. By Lemma \ref{GeneralRUM}, stochastic rationalizability is equivalent to there being a solution to Equations \ref{linconsist}-\ref{linnonneg}, and so we are done with this direction. We now show the other direction. Let $q$ be a solution to Equations \ref{linconsist}-\ref{linnonneg}. We proceed by induction on $A$.

The base case of our induction is when $A =X$. Note that either $X \in \mathcal{X}$ or $x \not \in \mathcal{X}$. If $X \not \in \mathcal{X}$, then Equations \ref{qcon}-\ref{qnonneg} coincide with Equations \ref{linconsist}-\ref{linnonneg} at $X$. If $X \in \mathcal{X}$, then we observe choice probabilities at $X$, and so $\tilde{q}(x,X) \geq 0$ for all $x \in X$ as $p(x,X)=\tilde{q}(x,X)$ by Equation \ref{qcon}. Since probabilities sum to one, this gives us $\sum_{x \in X} p(x,X)=\sum_{x\in X}\tilde{q}(x,X)=1$ which is exactly Equation \ref{lininitial}. Thus Equations \ref{linconsist}-\ref{linnonneg} hold at $X$. Now fix a set $A \subsetneq X$ and suppose that Equations \ref{linconsist}-\ref{linnonneg} hold for every set $B$ such that $A \subsetneq B$. This is our induction hypothesis. There are two cases; either $A \in \mathcal{X}$ or $A \not \in \mathcal{X}$. If $A \in \mathcal{X}$, then we know that $\sum_{x\in A}p(x,A)=1$ as choices are observed at $A$. Further, by our induction hypothesis $\sum_{B \subseteq B'}\tilde{q}(x,B')=p(x,B)$. It then follows that $\tilde{q}$ is the M\"{o}bius inverse of $p$ on the domain $\{B|A\subseteq B\}$. This gives us that $p(x,A)=\sum_{A \subseteq B}\tilde{q}(x,B)$. As $\tilde{q}(x,A)\geq 0$ is assumed in Equation \ref{qnonneg}, all we need to show is that Equation \ref{lininflow} holds at $A$. Since $p(\cdot)$ is set constant on $\{B|A\subseteq B\}$, it follows from Lemma \ref{qtop} that Equation \ref{lininflow} holds at $A$. Thus, if $A \in \mathcal{X}$, then Equations \ref{linconsist}-\ref{linnonneg} hold at set $A$. Now suppose that $A \not \in \mathcal{X}$. When $A \not \in \mathcal{X}$, Equations \ref{qcon}-\ref{qnonneg} coincide with Equations \ref{linconsist}-\ref{linnonneg} at set $A$. Thus by induction, if Equations \ref{qcon}-\ref{qnonneg} hold, then Equations \ref{linconsist}-\ref{linnonneg} hold.

\subsection{Proof of Proposition \ref{growthrate}}
We begin with the case that $\mathcal{X}=2^{X}\setminus \{\emptyset\}$. In this case, only Equation \ref{qcon} is encoded in our $N$ matrix. In this case, the matrix $N$ has one row for each pair $(x,A)$ with $x\in A \subseteq X$. This is given by $\sum_{i=1}^{|X|} i \binom{|X|}{i}=|X|2^{|X|-1}$. Now consider the general case. In the general case, for each $A \not \in \mathcal{X}$, we remove one case of Equation \ref{qcon} for each $x \in A$ and replace it with one case of Equation \ref{qinflowoutflow} (or Equation \ref{qinitialcon} if $A= X$). This change from the full domain case is equal to $|A|-1$, thus giving us $|X|2^{|X|-1}-\sum_{A \not \in \mathcal{X}}(|A|-1)$ and so we are done.

\subsection{Proof of Proposition \ref{monotoneequivalence}}
    From Theorem \ref{RUMEquivalenceThm}, we already know that stochastic rationalizability is equivalent to the existence of solutions to Equations \ref{qcon}-\ref{qnonneg}. All that is left to show is that Equation \ref{partial} is equivalent to rationalizability by monotone utilities in the case of stochastic rationalizability. By Theorem \ref{RUMmonotone}, we know that rationalizability by monotone utilities is equivalent to $p(x,A)=0$ whenever $x$ contains an alternative $y$ such that $y$ dominates $x$ in the underlying ordering of $X$. This is equivalent to $\sum_{A \subseteq A}q(x,B)=0$ for the chosen $(x,A)$. Further, note that if $A$ contains a $y$ dominating $x$, then every superset of $A$ contains a $y$ dominating $x$. Since we are in the case of stochastic rationalizability, this means that $q(x,B) \geq 0$ for all $(x,B)$. By the prior logic, this means $p(x,A)=0$ for $A$ with $y$ dominating $x$ is equivalent to $q(x,B)=0$ for all $A \subseteq B$. It then immediately follows that $q(x,B)=0$ for all $B$ with $y$ dominating $x$. In the case of $q(x,A) \geq 0$, this is equivalent to $\sum_{A:x \in A, A \cap U(x) \neq \emptyset}q(x,A)=0$ $\forall x \text{ such that }U(x) \neq \emptyset$, and so we are done.

\subsection{Proof of Theorem \ref{rumme}}
    Let $N(x,A) = \{\succ \in \mathcal{L}(X)|x \succ y \text{ } \forall y \in A \setminus \{x\}\}$. A random choice rule on $X$ is stochastically rationalizable if and only if there exists $\nu\in \Delta(\mathcal{L}(X))$ such that $p(x,A)=\sum_{\succ \in N(x,A)}\nu(\succ)$ for all $x\in A \in \mathcal{X}$. This is equivalent to the existence of such a $\nu$ and the existence of choice probabilities $p(y,B)$ for each $B \in (2^X \setminus\{\emptyset\})\setminus \mathcal{X}$ such that $p(x,A)=\sum_{\succ \in N(x,A)}\nu(\succ)$ for all $x\in A \in 2^X \setminus\{\emptyset\}$. By Theorem \ref{rumHrep}, this is equivalent to the existence of choice probabilities $p(y,B)$ for each $B \in (2^X \setminus\{\emptyset\})\setminus \mathcal{X}$ such that $q(x,A)\geq 0$ for each $x \in A \in 2^X \setminus\{\emptyset\}$. By Lemma \ref{qtop}, this is equivalent to the existence of $q(x,A)$ satisfying the conditions of Lemma \ref{GeneralRUM}.

    We now construct the matrix form of this linear program. Consider a matrix $D$ whose columns are indexed by $(x,A)$ for each $A \in 2^X \setminus\{\emptyset\}$ and each $x \in A$ and whose rows are indexed by $(y,B)$ for each $B \in \mathcal{X}$ and $y \in B$. The entry $d_{(y,B),(x,A)}=1$ if $B \subseteq A$ and $x=y$ and is equal to zero otherwise. $D$ encodes that our unobserved $q$ function must induce our observed choice probabilities. Let $P$ be a column vector indexed by $(y,B)$ for each $B \in \mathcal{X}$ and $y \in B$. Entry $p_{(y,B)}$ is equal to $p(y,B)$. Consider a matrix $E$ whose columns are indexed by $(x,A)$ for each $A \in 2^X \setminus\{\emptyset\}$ and each $x \in A$ and whose rows are indexed by $B \in 2^X \setminus\{X,\emptyset\}$. The entry $e_{B,(x,A)}$ is given as follows.
    \begin{equation*}
        e_{B,(x,A)}=\begin{cases} 1 & \text{ if } x \not \in B, A = B \cup \{x\} \\
        -1 & \text{ if } x \in B, A = B \\
        0 & \text{ otherwise}
        \end{cases}
    \end{equation*}
    $E$ encodes that our unobserved $q$ satisfy inflow equals outflow. Consider a row vector $F$ whose elements are indexed by $(x,A)$ for each $A \in 2^X \setminus\{\emptyset\}$ and each $x \in A$. The element $f_{(x,A)}$ is equal to one if $A=X$ and equal to zero otherwise. $F$ encodes that our unobserved $q$ satisfy $\sum_{x\in X}q(x,X)$. To be stochastically rationalizable, we must impose $q \geq 0$ which implies the last condition of Lemma \ref{qtop}. Thus the linear program we have constructed looks as follows.
    \begin{align}
        \left[\begin{array}{c}
        D\\
        \hline
        E\\
        \hline
        F
        \end{array}\right] q = \left[\begin{array}{c}
        P \\
        \hline
        \mathbf{0} \\
        \hline
        1
        \end{array}\right]
        \\
        q \geq 0
    \end{align}
    By Farkas's Lemma, (see Theorem 34 in \cite{borderAlternative} for a reference), there exists a solution to this linear program if and only if there does not exist a solution $r \in \mathbb{R}^M$ to the following linear program.
\begin{align}
    r^T\left[\begin{array}{c}
        D\\
        \hline
        E\\
        \hline
        F
    \end{array}\right] \leq 0 \label{alternativein}\\
    \left[\begin{array}{c|c|c}
        P & \mathbf{0} & 1
    \end{array}\right]\cdot r^T > 0 \label{alternativesum}
\end{align}
Writing out Equation \ref{alternativesum} gives us the following.
\begin{equation}\label{alternativesum2}
    \sum_{(y,B):y\in B \in \mathcal{X}} r(y,B)p(y,B) + r(X) > 0 
\end{equation}
Above, $r(X)$ is the element of $r$ which is associated with the vector $F$. As $r(X)$ can be any real number, we can rewrite Equation \ref{alternativesum2} as follows.
\begin{equation}\label{alternativesumrewrite}
    \sum_{(y,B):y\in B \in \mathcal{X}} r(y,B)p(y,B) - r(X) > 0
\end{equation}
For a given $(x,A)$, Equation \ref{alternativein} can be written as follows.
\begin{equation}\label{alternativeinre}
    \sum_{B:x \in B \subseteq A, B \in \mathcal{X}} r(x,B) + r(A \setminus \{x\}) - r(A) \leq 0
\end{equation}
Above, in the case that $A \setminus \{x\} = \emptyset$, we define $r(\emptyset)=0$. Further, $r(X)$ shows up as $-r(X)$ as a consequence of our transformation of Equation \ref{alternativesum2} into Equation \ref{alternativesumrewrite}. The negation of the existence of some $r$ solving Equations \ref{alternativesumrewrite} and \ref{alternativeinre} is equivalent to the negation of every locally feasible assignment and capacity pair $(a,c)$ being feasible. Thus it follows that the condition of Theorem \ref{rumme} holds if and only if $p$ is stochastically rational.

\bibliographystyle{ecta}
\bibliography{graph}

\end{document}